\documentstyle[12pt,fleqn,epsfig,axodraw]{article}

\newlength{\dinwidth}
\newlength{\dinmargin}
\newcommand{\tinyPomeron}{\mbox{\tiny $I\!\!P$}}

\newcommand{\lsim}{\raise.25ex \hbox{ $<$ \kern-1.1em \lower1ex \hbox {$\sim$ }}}
\newcommand{\gsim}{\raise.25ex \hbox{ $>$ \kern-1.1em \lower1ex \hbox {$\sim$ }}}

\newcommand{\vm}{vector meson}

\newcommand{\vmwf}{vector meson wave function}

\newcommand{\cs}{cross section} 
\newcommand{\css}{cross sections}

\newcommand{\qq}{\mbox{$q \overline{q}$}}

\newcommand{\qqpair}{\mbox{$q \overline{q} $-pair}}

\newcommand{\qql}{\mbox{$q \overline{q} $-loop}}

\newcommand{\etal}{{\it et al.}}
\newcommand{\ie}{{\it i.e.}}

\newcommand{\pert}{perturbative}
\newcommand{\nonpert}{non-perturbative}

\newcommand{\LandN}{Landshoff and Nachtmann}

\newcommand{\r}{\mbox{$\rho$}}

\newcommand{\pom}{pomeron}

\newcommand{\cm}{centre-of-mass}

\newcommand{\dep}{dependence}

\newcommand{\deps}{dependencies}

\newcommand{\qsq}{\mbox{$Q^{2}$}}

\newcommand{\gev}{{\rm Ge}\kern-1.pt{\rm V}}
\newcommand{\gevsq}{\mbox{$\mathrm{{\rm Ge}\kern-1.pt{\rm V}}^2$}}

\newcommand{\tinygev}{{\tiny \rm Ge}\kern-1.pt{\tiny \rm V}}
\newcommand{\tinygevsq}{\mbox{$\mathrm{{\tiny \rm Ge}\kern-1.pt{\tiny \rm V}}^2$}}

\newcommand{\kev}{{\rm ke}\kern-1.pt{\rm V}}
\newcommand{\kevsq}{\mbox{$\mathrm{{\rm ke}\kern-1.pt{\rm V}}^2$}}

\newcommand{\mev}{{\rm Me}\kern-1.pt{\rm V}}
\newcommand{\mevsq}{\mbox{$\mathrm{{\rm Me}\kern-1.pt{\rm V}}^2$}}

\newcommand{\mqsq}{\mbox{$m_{q}^{2}$}}

\newcommand{\beq}{\begin{equation}}
\newcommand{\eeq}{\end{equation}}

\newcommand{\beqarr}{\begin{eqnarray}}
\newcommand{\eeqarr}{\end{eqnarray}}

\newcommand{\aEM}{\mbox{$\alpha_{em}$}}

\newcommand{\aS}{\mbox{$\alpha_{\mbox{\tiny S}}$}}

\newcommand{\bfelltsq}{\mbox{\boldmath $\ell$}_t^2}

\newcommand{\bfktsq}{\mbox{\boldmath $k$}_t^2}

\newcommand{\bfktquad}{\mbox{\boldmath $k$}_t^4}

\newcommand{\Mxsq}{\mbox{$M_{X}^{2}$}}
\newcommand{\mxsq}{\mbox{$M_{X}^{2}$}}

\newcommand{\qbarsq}{\overline{Q}^2}

\begin{document}
\begin{flushright}
M/C-TH 00/07\\
September 2000\\
\end{flushright}
\begin{center}
\vspace*{2cm}

{\Large \bf 
Rho Electroproduction and the Hadronic Contribution to
Deeply Virtual Compton Scattering}

\vspace*{1cm}

A Donnachie, J Gravelis and G Shaw
\footnote{{\tt ad@theory.ph.man.ac.uk, janis@theory.ph.man.ac.uk,
graham.shaw@man.ac.uk}}

\vspace*{0.5cm}
Department of Physics and Astronomy, University of Manchester,\\
Manchester, M13 9PL, England.
\end{center}
\vspace*{2cm}

\begin{abstract}

A two-gluon-exchange model incorporating perturbative and
non-perturbative effects is presented for $\rho$ electroproduction
which provides an excellent description of all current data. This is
then used to calculate the contribution from the $\rho$ to deeply
virtual Compton scattering via the vector-meson-dominance transition
$\rho \to \gamma$.  This is found to be sufficiently large to provide
a significant contribution through interference with the perturbative
QCD term.

\end{abstract}


\newpage

\section{Introduction}

There is considerable evidence in $\gamma^{*} p$ reactions that the 
nominally perturbative regime can be strongly influenced by \nonpert\ 
effects.  This is an obvious feature of recent dipole models of deep 
inelastic scattering \cite{NNPZ97,FKS99,KM00}, where for transverse 
photons especially the contribution from large (non-perturbative) dipoles 
extends to significantly large values of $Q^2$. The penetration of 
non-perturbative physics into the perturbative regime is 
even more explicit in generalised vector dominance models \cite{KMS:95_97} 
or in two-component models~\cite{DL98,KM99,DDR99,DDR00,GLMN99,GLMN00} which
combine ``soft'' (non-perturbative) and ``hard'' (perturbative) 
contributions. Typically the soft contribution comprises the normal 
reggeon and soft pomeron exchanges, the latter with an intercept of 
$\sim 1.08$. The hard contribution may be a second pomeron, 
the hard pomeron, with an intercept of $\sim 1.44$ \cite{DL98,DDR99,DDR00}, 
or be based explicitly on perturbative QCD \cite{KM99,GLMN99,GLMN00}.

A good illustration of the two-component approach is provided by exclusive 
$\rho$ electroproduction, $\gamma^{*} p \to \rho p$. The high-energy data
\cite{{data:ZEUS_Breitweg:1999},{data:H1_Adloff:1999}} 
indicate that this approach is appropriate, as the effective pomeron
intercept increases from the canonical hadronic value of $\sim 1.08$
for real photons to perhaps as large as $\sim 1.19$ at $Q^2 = 20 \,
\gevsq$. In section 2 we present a two-component model for $\rho$
electroproduction which successfully describes all current data. We
base our calculations on two-gluon-exchange models of the pomeron. For
the non-perturbative contribution we use the model of
Diehl~\cite{MD:1995} and for the perturbative contribution that of
Martin, Ryskin and Teubner
\cite{Martin_Ryskin_Teubner:1996}. The procedure follows the suggestion of
\cite{Martin_Ryskin_Teubner:1996} by calculating the light quark 
anti-quark 
pair $u \overline{u}$ and $d \overline{d}$ production process $\gamma^{*} p
\to \qq \, p$ with the invariant mass of the \qqpair\ $M_X$ afterwards
integrated over the mass interval of the $\rho$. This approach has the 
benefit of avoiding  \vmwf\ complications, which can be serious~\cite{DGS00},
and allows one to concentrate on the production dynamics.

A related topic is that of Deeply Virtual Compton Scattering (DVCS) on 
protons, $\gamma^{*} p \to  \gamma p$, which is seen as an important 
reaction for the study of diffraction in QCD. In the standard QCD
approach the amplitude is described by skewed parton distributions \cite{G2}
corresponding to operator products evaluated between protons of
unequal momenta.  These are generalizations of the familiar parton
distribution of deep inelastic scattering, and like them satisfy
perturbative evolution equations \cite{G3} which enable them to be
evaluated at all \qsq\ in terms of an assumed input at some appropriate 
$Q^2 = Q_0^2$. Preliminary data \cite{H100} have been presented which 
are consistent with QCD predictions \cite{FFS}, subject to two
uncertainties.

The first is that the theoretical predictions refer to zero momentum
transfer $t = 0$ and to compare with experiment, one must integrate
over $t$. This is done by assuming an exponential dependence $
\exp(-bt)$ and estimating the unmeasured slope parameter $b$. The
considerable uncertainty\footnote{In analysing their preliminary data,
H1 \cite{H100} assume $ 7 \le b \le 10$ GeV$^{-2}$} in $b$ leads to a
corresponding uncertainty in the normalization of the predictions.

Secondly, it is necessary to specify the input skewed
parton distributions at the reference $Q_0^2$. In \cite{FFS} these are 
obtained by estimating their ratio to ``ordinary'' parton distributions at 
$Q_0^2 = 2.5 \, \gevsq$ using arguments based on the aligned jet model
\cite{AJM}, which in practice is almost identical to the simplest
diagonal  generalized vector meson dominance model for the
soft pomeron term \cite{VMD}. The resulting ratio is of the order of 2,
leading to a factor of order four in the predicted DVCS cross-sections. 
While this provides a reasonable first estimate, it is clearly subject 
to uncertainties which will become important when more accurate data 
are available.
 
Here we note that there are direct ``hadronic'' contributions to DVCS
via the vector-meson-dominance mechanism $\gamma^{*} p \to V p$, $V
\to \gamma$.  One particular vector-meson contribution to DVCS, namely
that of the $\rho$, can be calculated with reasonable precision using
the results of Section 2.  This is done in section 3, where we show
that the results provide useful constraints on models used to estimate
the skewed parton distributions at the reference $Q_0^2$.


\section{Rho electroproduction}

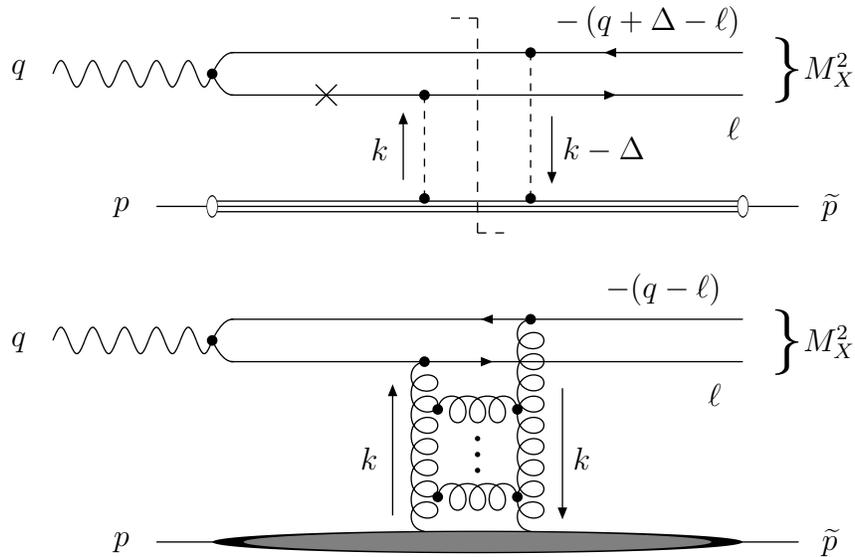
\begin{figure}[!t]
\begin{center} 
%
%
%
%
%
\begin{picture}(294,70)(20,90)
\Photon(30,140)(90,140){5}{5}
%
\Vertex(90,140){2}
\Curve{(90,140)(95,133)(98,132)}
\Curve{(90,140)(95,147)(98,148)}
%
\ArrowLine(290,148)(190,148)
\Line(190,148)(97,148)
\Line(97,132)(190,132)
\ArrowLine(190,132)(290,132)
%
\Line(129,128)(137,136)
\Line(129,136)(137,128)
%
\Line(91,88)(289,88)
\Line(91,90)(289,90)
\Line(91,92)(289,92)
\Line(69,90)(89,90)
\Line(291,90)(311,90)
%
\GOval(90,90)(4,2)(0){1}
\GOval(290,90)(4,2)(0){1}
%
\Vertex(170,93){2}
\Vertex(170,132){2}
\Vertex(210,93){2}
\Vertex(210,148){2}
\DashLine(170,93)(170,132){3}
\DashLine(210,93)(210,148){3}
%
\LongArrow(162,100)(162,124)
\Text(157,112)[cr]{$k$}
%
\LongArrow(218,124)(218,100)
\Text(223,112)[cl]{$k - \Delta$}
%
\DashLine(180,161)(190,161){4}
\DashLine(190,161)(190,80){4}
\DashLine(190,80)(200,80){4}
\Text(20,142)[r]{$q$}
\Text(59,90)[r]{$p$}
\Text(321,90)[l]{$\widetilde{p}$}
\Text(290,120)[cr]{$\ell$}
\Text(290,160)[cr]{$-\left( q + \Delta - \ell \right)$}
%
\Text(314,142)[r]{\Huge{\}}}
\Text(314,142)[cl]{$M_{X}^{2}$}
\end{picture}
%
%
\begin{picture}(294,125)(20,95)
\Photon(30,170)(90,170){5}{5}
%
\Vertex(90,170){2}
\Curve{(90,170)(95,163)(98,162)}
\Curve{(90,170)(95,177)(98,178)}
%
\ArrowLine(290,178)(97,178)
\ArrowLine(97,162)(290,162)
%
\Line(69,94)(110,94)
\Line(270,94)(311,94)
%
\GOval(190,94)(4,94)(0){0.5}
%
\Vertex(170,162){2}
\Vertex(210,178){2}
%
%
\Gluon(170,98)(170,162){5}{7}
\Gluon(210,98)(210,178){5}{9}
%
\Gluon(175,144)(205,144){5}{3}
\Gluon(175,111)(205,111){5}{3}
%
\Vertex(175,144){2}
\Vertex(205,144){2}
\Vertex(175,111){2}
\Vertex(205,111){2}
%
\Vertex(190,121){1}
\Vertex(190,127){1}
\Vertex(190,133){1}
%
\LongArrow(158,104)(158,152)
\Text(153,126)[cr]{$k$}
%
\LongArrow(222,152)(222,104)
\Text(227,126)[cl]{$k$}
\Text(20,170)[r]{$q$}
\Text(59,94)[r]{$p$}
\Text(321,94)[l]{$\widetilde{p}$}
\Text(283,150)[cr]{$\ell$}
\Text(283,190)[cr]{$-(q - \ell)$}
\Text(314,170)[r]{\Huge{ \}}}
\Text(314,170)[cl]{$M_{X}^{2}$}
\end{picture}
%
%
\end{center}
\caption{One of the four diagrams corresponding to the non-perturbative model 
\protect\cite{{MD:1995},{LN:1987}} (upper diagram), at $t = \Delta^2$, 
and the perturbative description
\protect\cite{Martin_Ryskin_Teubner:1996} (lower diagram), at $t = 0$.
The other three diagrams differ in the way gluon lines are attached to
the quarks in the \qql.
The minus sign in the four-momenta indicates an antiparticle.  
The off-shell quark and the cut along which the quark lines are put
on-shell in the LN model are indicated by the cross and the dashed
line respectively.
\label{open_production_diagrams}
}
\end{figure}

According to the factorisation theorem
\cite{factorisation_Collins_Frankfurt_Strikman:1997} 
the exclusive vector meson production processes can be factored into
three parts: the fluctuation of the (virtual) photon into a \qqpair;
the interaction of the \qqpair\ with the proton; and the formation of
the \vm\ from the \qqpair.  Similarly, the upper part of either
diagram in figure \ref{open_production_diagrams}, containing the
process $\gamma^{*} \to \qq$ can be considered separately from the
rest of the diagram. Apart from the couplings $\aS$, the upper parts
of the non-perturbative and perturbative diagrams are identical.  The
lower parts, that is the parts containing the gluons and the proton,
are described differently in each approach.

The model of Diehl~\cite{MD:1995} follows \LandN\ (LN) 
\cite{LN:1987}. The gluons are assumed not
to interact with each other and a \nonpert\ gluon propagator
\cite{MD:1995} is used:   
\beq
{\mathcal D}_{np} \left( - k^2 \right) \ = \ 
{\mathcal N}_{np} 
\left[ 1 + \frac{k^2}{\left( n - 1 \right) \mu_0^2 } \right]^{-n},
\label{eqn:nonpert_propagator}
\eeq
with $n = 4$. The normalisation ${\mathcal N}_{np}$ is determined from
the condition
\beq
\int_{0}^{\infty} dk^{2} \left[ \alpha_{S}^{(0)} 
{\mathcal D}_{np}(k^{2}) \right]^{2} 
\ = \ \frac{9 \beta_{0}^{2}}{4 \pi} ~.
\label{eqn:ln_prop_normalisation_1}
\eeq
The phenomenological parameters $\beta_{0}$,
which describes the effective coupling of the pomeron to the proton, 
and $\mu_{0}$ are determined
from the total $pp$ and $p \overline{p}$ \cs\ data and from deep inelastic 
scattering: $\beta_{0} \approx 2.0 \ \gev^{-1}$ and $\mu_{0} \approx
1.1 \ \gev$ \cite{DL:1988_1989}. For the non-perturbative couplings
of the gluons to the quarks forming the $\rho$ a value 
$\alpha_{S}^{(0)} \approx 1$ is taken.

It has been argued \cite{LN:1987} that the diagrams in which the
non-perturbative gluons couple to different valence quarks in the
proton are suppressed and therefore can be disregarded.  Only the
diagrams where both gluons couple to the same valence quark are
calculated.  Each of the three valence quarks is incorporated into the
proton according to the Dirac form factor of the proton $F_{1p}(t)$,
where $t = \Delta^2$ and the four-momenta of the particles are as
depicted in figure \ref{open_production_diagrams}. The energy \dep\
due to the soft \pom\ comes via a factor
$x_{\tinyPomeron}^{-\alpha_{\tinyPomeron}(t)}$ in the amplitude, with
$x_{\tinyPomeron} \equiv (M_X^{2} + Q^{2} - t)/ (W^{2}+Q^{2}-
m_{proton}^{2})$ and $\alpha_{\tinyPomeron}(t)$ the soft
\pom\ trajectory \cite{DL:1986_pomeron_intercept}. In principle we can 
calculate the $t$-dependence of the soft-pomeron contribution from this,
but as this cannot be done for the perturbative contribution, we 
calculate only at $t = 0$ and use the experimental slope to give the
integrated cross section.

Following the argument of \cite{MD:1995}, the coupling $\aS$ at three 
of the four vertices is taken at a \nonpert\ scale, \ie\
$\alpha_S^{(0)}$ is used, while for the vertex where the gluon couples
to the off-shell quark it is taken at a \pert\ scale
$\lambda^2 = \left( \bfelltsq + \mqsq \right) \left(
\qsq + \mxsq \right) / \mxsq$, which is a typical
scale for the whole upper part of the diagram.

In the \pert\ approach by Martin \etal\ \cite{Martin_Ryskin_Teubner:1996}
the \pom\ is modelled as a
pair of perturbative gluons with symmetric momenta. The perturbative 
gluon propagator ${\mathcal D}_{p}(k^{2}) = 1 / k^{2}$ is used. 
In principle the
gluon flux can be obtained from the unintegrated gluon density 
$f (x_{\tinyPomeron}, |k^2|)$, which
gives the probability of finding a $t$-channel gluon with the
momentum squared $|k^2|$ in the proton. However, a
special treatment of the infrared region is required as the unintegrated
gluon density $f (x_{\tinyPomeron}, |k^2|)$ is theoretically undefined
as $|k^2| \to 0$ and numerically unavailable below some value of
$|k^2| = Q_0^2$, which varies with the parton distribution chosen and
usually is in the region from 0.2 to a few \gevsq.
The linear approximation as suggested in
\cite{Martin_Ryskin_Teubner:1996} is used to account for the 
contribution to the integral from the $|k^2| < Q_0^2$ region.
A number of gluon distributions were tried, including MRS(R1), MRS(R2),
GRV94HO, GRV94LO and others using the PDFLIB program libraries
\cite{PDFLIB_manual} for numerical calculations.
However, the one which gave the best energy dependence within the
model is CTEQ4LQ \cite{CTEQ}, so results are presented only for that
choice.

The derivations of both models can be expressed in a common kinematical
framework, taking into account the on-shell conditions along the cut
line in figure \ref{open_production_diagrams}, which result in the gluon
momenta being predomimently transverse  with respect 
to the $\gamma^{*} p$ axis, $|k^2| \approx \bfktsq$. Here and subsequently 
transverse two-vectors are shown in bold.
Integrating over the azimuthal angles, one obtains a
common structure for both models~\cite{JG_thesis:2000} at $t = 0$:
\beq
\frac{d^2 \sigma^{L, \, Tr}}{d{\mxsq} dt} \ = \
\frac{16 {\mathrm e}^2_q \aEM }{3} 
\frac{1}{\mxsq}
\int_{0}^{\frac{1}{4} M_{X}^{2} - m_{q}^{2}} 
\frac{d \bfelltsq}{\sqrt{1 - 4 ( \bfelltsq + \mqsq ) / \mxsq }} 
\left( \frac{\bfelltsq + \mqsq}{\mxsq} \right)
\ {\mathcal S}^{L, \, Tr}
\label{eqn:MD_cs_tr}
\nonumber
\eeq
with
\beqarr
{\mathcal S}^{L} &=& 
4 \qsq \left( \frac{\bfelltsq + \mqsq}{\mxsq} \right)^2
\left[
\int \! d \bfktsq  
\ {\mathcal P} \
\left(
\frac{1}{\qbarsq + \bfelltsq }
\: - \: \frac{1}
{\sqrt{( \qbarsq + \bfelltsq + \bfktsq )^2
\: - \: 4 \bfktsq \bfelltsq } } \right)
\right]^2
\label{eqn:md_S_L}
\\
{\mathcal S}^{Tr} &=&
\left[
\int \! d \bfktsq  
\ {\mathcal P} \
\left( \frac{1}{\overline{Q}^2 + \bfelltsq} 
\: - \: \frac{1}{2 \bfelltsq} \: + \: \frac{\overline{Q}^2 - \bfelltsq +
\bfktsq}{2 \bfelltsq 
\sqrt{( \qbarsq + \bfelltsq + \bfktsq )^2
\: - \: 4 \bfktsq \bfelltsq } } \right)
\right]^2 
\nonumber
\\
&\times& 
\bfelltsq 
\left( 1 - \frac{2 ( \mqsq + \bfelltsq )}{\mxsq} \right) 
\ + \ \frac{1}{4 Q^2} \left( \frac{\mxsq}{\bfelltsq + \mqsq} \right)^2
\mqsq \ {\mathcal S}^{L} 
\label{eqn:md_S_T}
\eeqarr 
where $\qbarsq \equiv \mqsq + Q^2 \left( \bfelltsq + \mqsq \right) / \mxsq$ 
and the symbol $\mathcal P$ denotes the model-dependent parts
\begin{eqnarray}
{\mathcal P}_{np} &=&
F_{1p}(0) \ 
x_{\tinyPomeron}^{ 1 - \alpha_{\tinyPomeron} \left( 0 \right) } 
\left[ {\alpha}_{\mbox{\tiny S}}^{(0)} \right]^{3/2} 
\sqrt{\aS \! \left( \lambda^2 \right)} 
\ \left[ {\mathcal D}_{np} \left( - k^2 \right) \right]^2 
\label{eqn:open_nonpert_prop_related_expr}
\\
\nonumber
\\
{\mathcal P}_{p} &=&
\frac{\pi}{4} \: \aS \! \left( \bfktsq \right) \:
f \! \left( x_{\tinyPomeron}, \bfktsq \right) \:
\frac{1}{\bfktquad}~.
\label{eqn:open_pert_prop_related_expr}
\end{eqnarray}
Here $f( x_{\tinyPomeron}, \bfktsq)$ is related to the gluon distribution
$g(x_{\tinyPomeron},Q^2)$ by 
\beq
x_{\tinyPomeron} \, g(x_{\tinyPomeron},Q^2) \ = \ 
\int^{Q^2}{{d\bfktsq}\over{\bfktsq}}
\, f( x_{\tinyPomeron}, \bfktsq)
\label{ftog}
\eeq
with the inverse
\beq
f( x_{\tinyPomeron}, \bfktsq) \ = \ 
\bfktsq \, {{\partial\Big(x_{\tinyPomeron}, 
g(x_{\tinyPomeron},\bfktsq)\Big)}\over{\partial\bfktsq}}~.
\label{gtof}
\eeq
Such a common structure should be present since both models describe
the same physical process.  The common parts originate mainly
from the kinematics of the process. The difference is contained in
equations (6) and (7), arising from the different physical interpretations 
of the internal dynamics of the process in the two models.

The relation between the models can formally be written as a
replacement ${\mathcal P}_{np} \longleftrightarrow {\mathcal P}_{p}$.
There are no $\sqrt{\aS}$ couplings for the two bottom vertices in the
\pert\ case since the gluons are considered as part of the proton and 
described by $f \! \left( x_{\tinyPomeron}, \bfktsq \right)$.
A minor difference is the different argument of $\aS$ in both models
and the fact that there is no need for the linear approximation in the
nonperturbative model since the integration over the gluon momentum can be
performed down to zero.
The expressions above are given in a general form but can be further
simplified for the light quarks assuming $m_q = 0$.

\begin{figure}[!p]
\vspace*{-5mm}
\begin{minipage}[t]{0.49\textwidth}
\begin{center}
\psfig{file=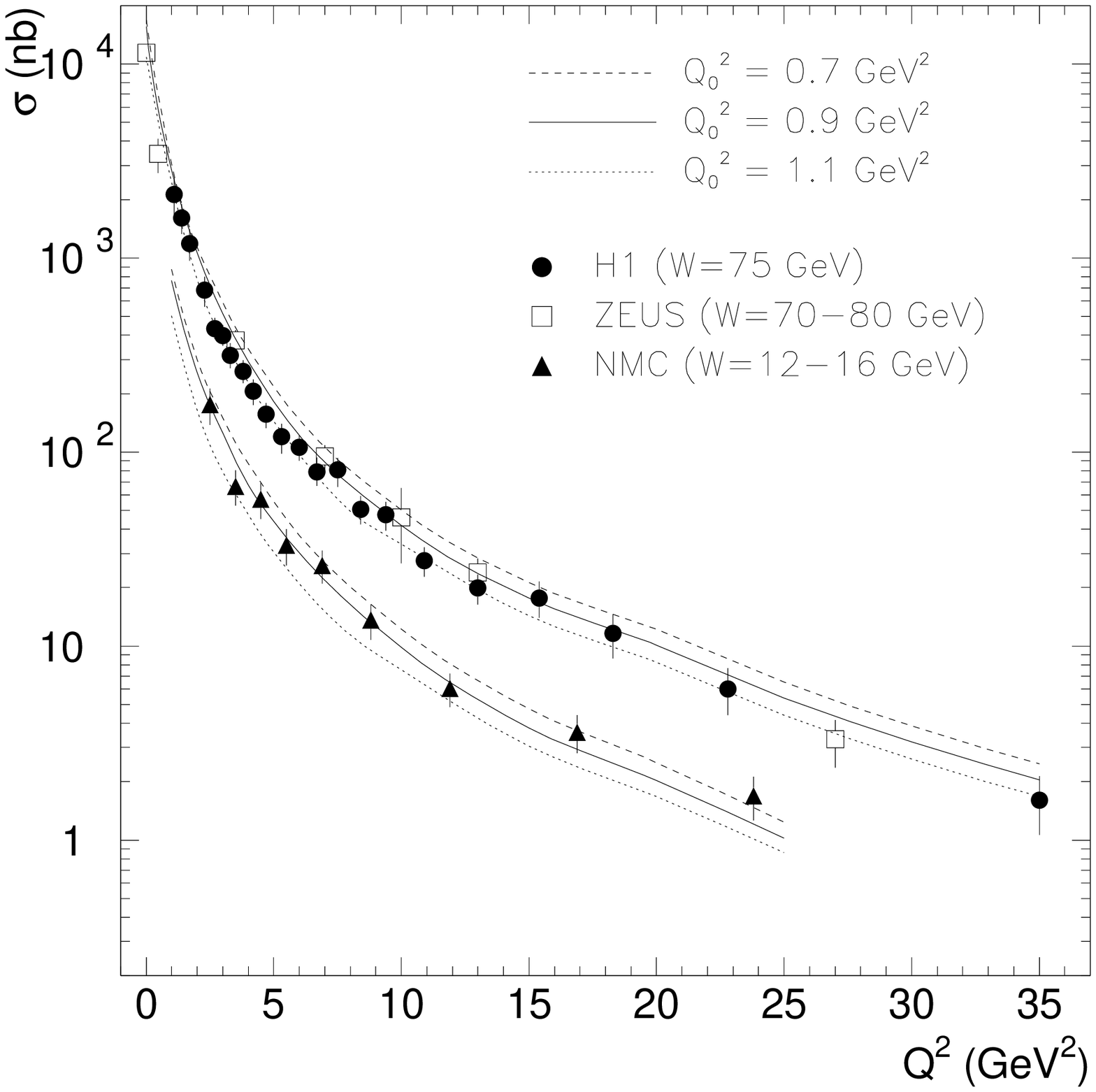, width=\textwidth}
\vspace*{-10mm}
\caption{ 
\label{fig:total_cs_Q0}
The \dep\ of total \cs\ on $Q_0^2$ compared with high energy data from
H1 \protect\cite{data:H1_Adloff:1999} and ZEUS
\protect\cite{{data:ZEUS_Breitweg:1999},{data:ZEUS_Breitweg:1998},{data:ZEUS_Derrick:1995}},
and low-energy data from NMC \protect\cite{data:NMC:1994}.
}
\end{center}
\end{minipage}
\hfill
\begin{minipage}[t]{0.49\textwidth}
\begin{center}
\psfig{file=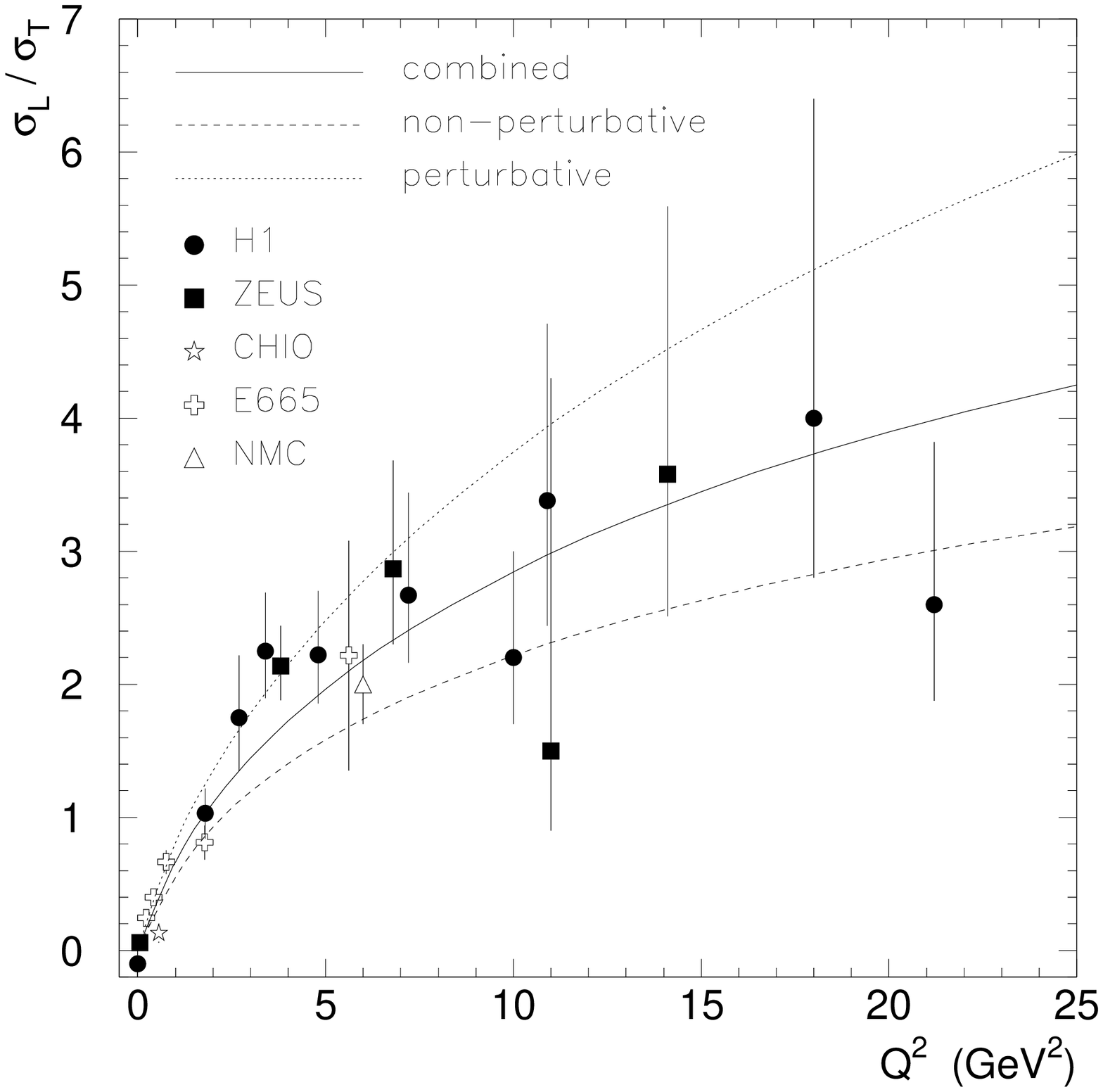, width=\textwidth}
\vspace*{-10mm}
\caption{ 
\label{fig:long_tr_ratio}
The \qsq-\dep\ of the longitudinal to transverse ratio compared with
high energy data from H1
\protect\cite{{data:H1_Adloff:1999},{data:H1_Aid:1996_1},{data:H1_Aid:1996_2}} and
ZEUS
\protect\cite{{data:ZEUS_Breitweg:1999},{data:ZEUS_Derrick:1995},{data:ZEUS_Derrick:1995_2}},
and low energy data from CHIO
\protect\cite{data:CHIO_Shambroom:1982}, E665
\protect\cite{data:E665_Adams:1997} and NMC \protect\cite{data:NMC:1994}.
}
\end{center}
\end{minipage}
\end{figure}
%
%
\begin{figure}[!p]
\vspace*{-30mm}
\begin{minipage}[t]{0.49\textwidth}
\begin{center}
\psfig{file=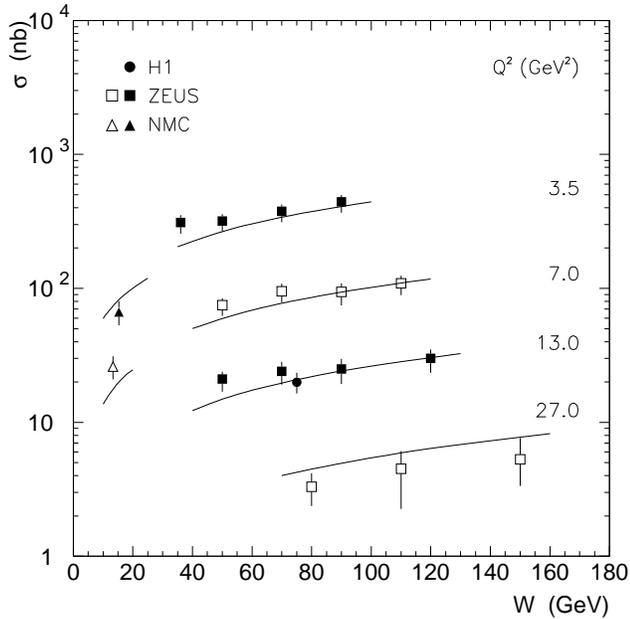, width=\textwidth}
\end{center}
\end{minipage}
\hfill
\begin{minipage}[t]{0.49\textwidth}
\begin{center}
\psfig{file=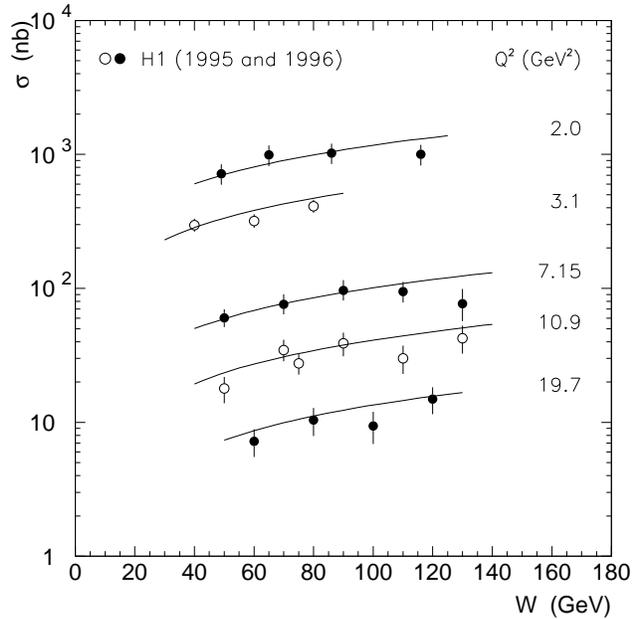, width=\textwidth}
\end{center}
\end{minipage}
\vspace*{-2mm}
\caption{ 
\label{fig:W_dep}
$W$-\dep\ compared with data from H1 \protect\cite{data:H1_Adloff:1999}, ZEUS
\protect\cite{data:ZEUS_Breitweg:1999} and NMC \protect\cite{data:NMC:1994}.
The use of the open and filled markers is only to indicate to which value
of \qsq\ the data points belong.
}
\end{figure}
%

The forward differential \cs\ $d \sigma / dt |_{t=0}$ can be related 
to the total \cs\ $\sigma (W, Q^2)$ using the experimentally measured 
forward diffractive slope $b_{\rho}(Q^2)$ assuming an exponential 
$t$-\dep\ of $d \sigma / dt$:
\beq
\sigma_{\gamma^{*} p \to \rho p} \ \simeq \ 
\frac{1}{b_{\rho}(Q^2)} 
\sum_{q = u,d} \: 
\int_{M_1^2}^{M_2^2} 
d\Mxsq \
\left[
\varepsilon_{expt} \, 
\frac{d\sigma^{L}_{\gamma^{*} p \to q \overline{q} \, p}}{dt \: d\Mxsq}
\ + \
\frac{d\sigma^{Tr}_{\gamma^{*} p \to q \overline{q} \, p}}{dt \: d\Mxsq}
\right]_{t=0}~.
\label{eqn:open_Mxsq_integration_1}
\eeq
The slope parameter $b_\rho(Q^2)$ varied from 7 at the smallest $Q^2$ to
4 at the highest $Q^2$.
The $M_X^2$-integration limits  $M_1^2 = \left( 0.6 \, 
\gev \right)^2$ and $M_2^2 = \left( 1.05 \, \gev \right)^2$ were chosen to
span the $\rho$-region, following
\cite{Martin_Ryskin_Teubner:1996}.  The polarisation of the photon
beam $\varepsilon_{expt}$ is a known characteristic of the experiment.
For HERA $\varepsilon_{expt} \approx 1$ while for fixed-target experiments 
it varies significantly, depending on the energy and photon virtuality. 
It is generally in the range of 0.5 to 0.9 and this is taken into account 
when comparing our results with the data.

Neither model by itself can describe all the observed features of the
\r\ electroproduction data simultaneously: that is, the absolute value
of the total \cs\ $\sigma(W, Q^2)$; its \deps\ on \qsq\ and $W$; and
the variation of the longitudinal to transverse \cs\ ratio $R =
\sigma^L / \sigma^T$ with \qsq. 
The \nonpert\ approach gives an energy dependence which is too flat at
the higher values of $Q^2$ due to the soft
\pom\ intercept while the energy \dep\ of the \pert\ approach, 
coming from the energy \dep\ of the gluon density, is clearly too
steep at all $Q^2$.  The \pert\ model does not replicate the
\qsq-\dep\ of the NMC data, and the non-perturbative model gives a
longitudinal to transverse ratio which is somewhat low\footnote{For further
details, see~\cite{JG_thesis:2000}}. 

The results
of summing the \pert\ and \nonpert\ production amplitudes
\beq
{\mathcal P} \ = \ 
{\mathcal P}_{np} \ + \ {\mathcal P}_{p}
\label{eqn:open_nonpert_sum_replacement}
\eeq
with ${\mathcal P}_{np}$ and ${\mathcal P}_{p}$ as given by
(\ref{eqn:open_nonpert_prop_related_expr}) and  
(\ref{eqn:open_pert_prop_related_expr}),
are shown in figures 
\ref{fig:total_cs_Q0}, \ref{fig:long_tr_ratio} and \ref{fig:W_dep}
together with the data. It is clear that the two-component model 
gives excellent agreement.
The effect of the value of $Q_0^2$ is shown in
figure \ref{fig:total_cs_Q0}.  The effect is consistent both at high and low
energy and allows us to fix $Q_0^2 \approx 0.9 \ \gevsq$, which 
is a reasonable value.


\section{Deeply virtual Compton scattering (DVCS)}

We  now use the results of the previous section to estimate the
contribution to DVCS from the mechanism of figure \ref{VMD}; and then comment
on its implications for the estimation of the skewed parton distributions at
the reference $Q_0^2 = 2.6 \, \gevsq$.

\subsection{Estimating the $\rho$ contribution}

\begin{figure}[!t]
\begin{center} 
%
%
\begin{picture}(175,75)(0,0)
%
\ArrowLine(0,-5)(80,25)
\ArrowLine(80,25)(160,-5)
\Photon(0,75)(80,25){4}{10}
\Line(80,25)(125,51)
\Line(80,28)(125,54)
\Vertex(125,52){3}
\Photon(125,52)(160,75){4}{5}
\GOval(80,25)(15,20)(0){0.5}
\Text(-5,0)[br]{$p$}
\Text(168,0)[bl]{$\widetilde{p}$}
\Text(-5,75)[cr]{$\gamma^{*}$}
\Text(168,75)[cl]{$\gamma$}
\Text(105,52)[bc]{$\rho$}
\end{picture}
\end{center}
%
%
%
\caption{
The hadronic contribution to DVCS arising from a virtual $\r$ in the final
 state. 
%
%
\label{VMD}
}
\end{figure}
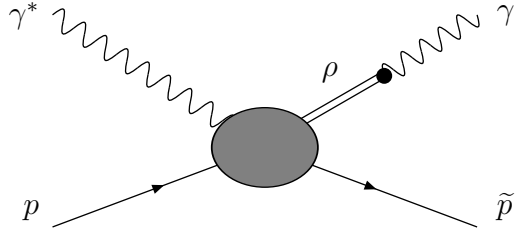

Assuming $s$-channel helicity conservation, only transverse photons 
contribute to the DVCS \cs.  
The relation between the DVCS and the $ep \to e \gamma p$ \css\ 
is~\cite{ZEUS:kin_relation:1994}
\beq
\frac{d^2 \sigma_{ep \to e \gamma p}}{dW \ d Q^2} \ = \
\frac{\aEM}{\pi} \, \frac{W}{Q^2 \left( W^2 + Q^2 - m_{proton}^2 \right)}
\left[ 1 + \left( 1 - y \right)^2 \right] \ 
\sigma^{Tr}_{\gamma^{*} p \to \gamma p}~,
\label{eqn:ep_gamma_p_kin_relation}
\eeq
with $y \equiv \left( W^2 + Q^2 - m_{proton}^2 \right) / \left( s -
m_{proton}^2 \right)$, $\sqrt{s}$ is the \cm\
energy of the $ep$ system and $W$ is that of the $\gamma^{*} p$ system.
The contribution from the mechanism of figure \ref{VMD} 
is then given by
\beq
\sigma_{\rho}(\gamma^* p \to \gamma p) \ \approx \ {{4\pi\alpha}\over
{\gamma^2_\rho}} \ \sigma(\gamma^* p \to \rho p)~.
\label{eqn:DVCS_cross_section}
\eeq
The coupling $e/\gamma_\rho$ of the $\rho$ to the photon is directly
related to the width of the decay $\rho \to e^+e^-$. In the
narrow-width approximation
\beq
\frac{4 \pi}{\gamma_{\rho}^{2}} \ = \ 
\frac{3 \ \Gamma_{\rho \to e^+ e^-}}{m_{\rho} \, \alpha^{2}}
\ = \ 0.494 \pm 0.023
\label{eqn:coupling}
\eeq
where the experimental values of $m_{\rho}$, $\Gamma_{\rho}$ and the
branching ratios have been used \cite{PDG}.

The values obtained from (\ref{eqn:ep_gamma_p_kin_relation}),
(\ref{eqn:DVCS_cross_section}) and (\ref{eqn:coupling}) for the
differential \cs\ $d \sigma_{\rho} / d Q^2$, averaged over each
\qsq-bin and integrated over $30 \ \gev < W < 120 \ \gev$ are shown in
Table 1, together with the preliminary experimental values
\cite{H100} for $d \sigma / d Q^2$ with the Bethe-Heitler term
subtracted off. As calculated, the amplitude in the model is purely
imaginary\footnote{There is no interference term, since the data are
integrated over the azimuthal angle $\phi_r$. (See eqn.(40) of
\cite{FFS})}.  We also give the estimate
\beq
R_{\mathcal A} \left( Q^2 \right) \ = \ 
\frac{{\mathcal A}_{\rho} \left( \gamma^{*} p \to \gamma p \right)}
{{\mathcal A} \left( \gamma^{*} p \to \gamma p \right)} 
\ \approx \
\left[ \frac{ d \sigma_{\rho} / d Q^2}{d \sigma / d Q^2} \right]^{1/2}
\eeq
of the amplitude of figure \ref{VMD} to the total non-Bethe-Heitler amplitude. 
As can be seen, this reduces from of order 20 $\%$ in the lowest $Q^2$
bin to of order 10 $\%$ in the highest. Although small, this is not
a negligible contribution to the total amplitude for DVCS, since it will
interfere constructively with the remaining dominant contributions assuming
they are mainly imaginary. It is also worth noting that even for the lowest 
$Q^2$ bin, ${{\mathcal A}_{\rho} \left( \gamma^{*} p \to \gamma p \right)}$,
which one might expect to be predominantly ``soft'', contains a significant 
contribution, about 40 $\%$, from the perturbative term.

\begin{center}
\begin{tabular}{c|c|c|c}
\hline\hline
$30 \, \gev < W \! < 120 \, \gev$ & $d \sigma_{\rho} / d Q^2$ & $d \sigma_{expt} / d Q^2$ &
$R_{\mathcal A}$ \\
$Q^2$ bin $\left[ \gevsq \right]$ & $\left[ pb \, / \gevsq \right]$ & 
$\left[ pb\, / \gevsq \right]$ & \\
%
\hline
2.0 to 4.0 & 1.576 & $47 ^{+12}_{-10}$ & $0.18^{+0.02}_{-0.02}$ \\
4.0 to 6.5 & 0.213 & $6.5 ^{+1.6}_{-2.5}$ & $0.18^{+0.04}_{-0.02}$ \\
$\,$ 6.5 to 11.0 & 0.0352 & $2.10 ^{+0.51}_{-0.64}$ & $0.13^{+0.02}_{-0.02}$ \\
11.0 to 20.0 & 0.00498 & $0.35 ^{+0.17}_{-0.14}$ & $0.12^{+0.02}_{-0.03}$ \\
\hline\hline
\end{tabular}
\begin{table}[!!!h]
\caption{Comparison of the preliminary experimental values of $d \sigma / 
d Q^2$ after subtraction of the Bethe-Heitler contribution, integrated over
$30 \, \gev < W \! < 120 \, \gev$, with the hadronic contribution of
figure \ref{VMD}.  The corresponding ratio of amplitudes $R_{\mathcal
A}$ is also given, assuming the amplitudes are pure imaginary.
\label{table:cross_sections}
}
\end{table}
\end{center}
%


\subsection{Estimating the input distributions}

In \cite{FFS} the input skewed parton distributions at the reference
$Q_0^2 = 2.6 \, \gevsq$ were estimated by relating them to
``ordinary'' parton distributions by using arguments based on the
aligned jet model \cite{AJM}. In practice this is almost identical to
the application of generalised vector meson dominance (GVD) in its
diagonal form \cite{VMD}.  Both models assume that for the scattering
of virtual photons the amplitude at $t = 0$ is of the form
\beq
{\rm Im} \, {\mathcal A} \, (\gamma^* N \to \gamma^* N)_{t=0} 
\ = \ {{\alpha}\over{3\pi}}
\int_{m_0^2}^{\infty}dm^2 \, {{m^2\rho(m^2)}\over{(Q^2+m^2)^2}}
\label{amp}
\eeq
where $\rho(m^2)$ is taken to be energy independent, corresponding to 
soft Pomeron behaviour with intercept $\alpha_{\tinyPomeron} = 1$.
For the aligned jet model $\rho(m^2)$ is given by
\beq
\rho(m^2) \ = \ \sigma^{\rm Tot}_{AJM}(m^2) \, R^{e^+e^-}(m^2) \,
{{3\langle k^2_T \rangle}\over{m^2}}~.
\label{ajm}
\eeq
In \cite{FFS} the product $ \sigma^{\rm Tot}_{AJM} \langle k^2_T \rangle
R^{e^+e^-}(m^2)$ is assumed to
be a constant. For diagonal vector meson dominance $\rho(m^2)$ is given by
\beq
\rho(m^2) \ = \ \sigma^{\rm Tot}_{Vp}(m^2) \, R^{e^+e^-}(m^2)~,
\label{vmd}
\eeq
and the  identification of the two approximations is completed by the usual 
diagonal GVD assumption that $\sigma^{\rm Tot}_{Vp}(m^2) \sim 1/m^2$ for  
$\alpha_{\tinyPomeron} = 1$ and $R^{e^+e^-}(m^2)$ is constant.

In the case of DVCS the imaginary part of the amplitude for $t = 0$ is
obtained from (\ref{amp}) by replacing one of the propagators with
$1/m^2$. Since $m^2\rho(m^2)$ 
in (\ref{amp}) and in its DVCS equivalent is taken to be  constant in  
the aligned jet/GVD model,  the integrals are trivial,  giving the result
\beq
{{{\rm Im} \, {\mathcal A} (\gamma^* N \to \gamma N)_{t=0}}
\over{{\rm Im} \, {\mathcal A} (\gamma^* N \to \gamma^* N)_{t=0}}} 
\ = \ {{1}\over{Q^2}} \, (m_0^2+Q^2) \, \ln(1+Q^2/m_0^2)~.
\label{ratio}
\eeq
For a reasonable choice of the lower limit $m_0^2$ in (\ref{ratio}),
typically 0.4 to 0.6 \gevsq, the ratio $\sim 2$ for $Q^2 \approx 2.5
\, \gevsq$. So knowing $F_2(x,Q^2)$ this gives the input to the DVCS
evolution equations \cite{FFS}.

At this point we notice that the same aligned jet/GVD model also
implies values for the $\rho$ contribution to DVCS. Attributing the
low mass contribution to the $\rho$-meson, one easily obtains
\beq
R_A(Q^2) \ = \
{{{\rm Im} \, A_{\rho}(\gamma^* N \to \gamma N)_{t=0}}\over{{\rm Im} \, 
A(\gamma^* N \to \gamma N)_{t=0}}} \ = \ \frac{\ln(1+Q^2/M_1^2) 
\, - \, \ln(1+Q^2/M_2^2) }{ \ln(1+Q^2/M_1^2)}
\label{ratio2}
\eeq
where we have chosen $m_0^2 = M_1^2 = ( 0.6 \, \gev )^2$ and $M_2^2 =
(1.05 \, \gev)^2$ for consistency with our treatment of the $\rho$ in
Section 2 (cf. equation \ref{eqn:open_Mxsq_integration_1}).  This
gives values for $R_{\mathcal A}$ of 0.43 at $Q^2 = Q^2_0 = 2.6 \,
\gevsq$ reducing to 0.28 at $Q^2 = 15 \, \gevsq$. These values are
considerably larger than the much more reliable estimates of section
3.1.

This discrepancy is not surprising given the extreme simplicity of the
aligned jet/GVD model used. More elaborate versions can be constructed
and will be required in the future to estimate more accurately the
skewed parton distributions at the reference $Q_0^2$. Such models will
need to take account of both ``hard'' and ``soft'' diffraction, and in
constructing them the requirements that they be compatible with the
$\rho$ contributions of Section 3.1, as well as the structure function
data for $ 0 \le Q^2 \le Q_0^2$, will be essential constraints.


\section*{Acknowledgments}

This work was supported by an Overseas Research Student Award, a
University of Manchester Research Scholarship and by PPARC grant
number PPA/G/0/1998. We thank Marcus Diehl, Martin McDermott and 
Thomas Teubner for many helpful comments.



\end{document}